\documentclass[conference]{IEEEtran}
\IEEEoverridecommandlockouts
\usepackage{underscore}
\usepackage{hyperref}
\usepackage{cite}
\usepackage{amsmath,amssymb,amsfonts}
\usepackage{algorithmic}
\usepackage{graphicx}
\usepackage{textcomp}
\usepackage{multirow}
\usepackage{xcolor}
\def\BibTeX{{\rm B\kern-.05em{\sc i\kern-.025em b}\kern-.08em
    T\kern-.1667em\lower.7ex\hbox{E}\kern-.125emX}}
\begin{document}
\title{Mixture of Experts Fusion for Fake Audio Detection Using Frozen wav2vec 2.0 \\
}


\author{
    Zhiyong Wang\textsuperscript{1,2},
    Ruibo Fu\textsuperscript{2,\textsuperscript{\dag}}, 
    Zhengqi Wen\textsuperscript{3},
    Jianhua Tao\textsuperscript{3},
    Xiaopeng Wang\textsuperscript{1,2}, 
    Yuankun Xie\textsuperscript{2},\\
    Xin Qi\textsuperscript{1,2},  
    Shuchen Shi\textsuperscript{2},
    Yi Lu\textsuperscript{1,2},
    Yukun Liu\textsuperscript{1},
    Chenxing Li\textsuperscript{4},
    Xuefei Liu\textsuperscript{2},
    Guanjun Li\textsuperscript{2},
    \vspace{1.5mm}
    \\
    \IEEEauthorblockA{\textsuperscript{1}University of Chinese Academy of Sciences, Beijing, China}
    \IEEEauthorblockA{\textsuperscript{2}Institute of Automation, Chinese Academy of Science, Beijing, China}
    \IEEEauthorblockA{\textsuperscript{3}Beijing National Research Center for Information Science and Technology, Tsinghua University, Beijing, China}
    \IEEEauthorblockA{\textsuperscript{4}AI Lab, Tencent, Beijing, China}
    \IEEEauthorblockA{wangzhiyong22@mails.ucas.ac.cn, ruibo.fu@nlpr.ia.ac.cn.}
    \thanks{\textsuperscript{\dag} Corresponding author.  }
    \thanks{Code is released at \href{https://github.com/john852517791/pytorch_lightning_FAD}{github.com/john852517791/pytorch_lightning_FAD}.
}
    
}

\maketitle

\begin{abstract}
 Speech synthesis technology has posed a serious threat to speaker verification systems. 
 Currently, the most effective fake audio detection methods utilize pretrained models, and integrating features from various layers of pretrained model further enhances detection performance. 
 However, most of the previously proposed fusion methods require fine-tuning the pretrained models, resulting in excessively long training times and hindering model iteration when facing new speech synthesis technology. 
 To address this issue, this paper proposes a feature fusion method based on the Mixture of Experts, which extracts and integrates features relevant to fake audio detection from layer features, guided by a gating network based on the last layer feature, while freezing the pretrained model. 
 Experiments conducted on the ASVspoof2019 and ASVspoof2021 datasets demonstrate that the proposed method achieves competitive performance compared to those requiring fine-tuning.
\end{abstract}

\begin{IEEEkeywords}
Fake Audio Detection, Mixture of Experts, Pretrained Model, Wav2Vec 2.0, Audio DeepFake Detection
\end{IEEEkeywords}

\section{Introduction}
The rapid advancement of speech synthesis and voice conversion technologies has enabled the creation of highly realistic fake audio at a low cost, posing significant challenges in distinguishing it from genuine audio. This capability raises concerns about the misuse of such technology for spreading misinformation or compromising automatic speaker verification systems. \cite{clsurvey}.
To prevent the misuse of fake audio and the resulting negative consequences, the research community proposes various Fake Audio Detection (FAD) methods. 
These methods can be categorized into two types based on whether they use pretrained model. 
Models that do not rely on pretrained systems are typically referred to as 'small models' due to their reduced number of parameters.
Some of these models that use handcrafted features, such as Short-Time Fourier Transform \cite{stft} and Constant Q Cepstral Coefficients \cite{cqcc}, combined with neural network classifier for binary classification. 
In addition to models based on handcrafted features, there are also end-to-end models, such as the Rawnet2 \cite{rawnet2}, TSSD \cite{tssd}, and AASIST \cite{aasist}, which take waveform input directly.
While these small models achieve good detection performance, they generally suffer from poor generalization, which means they often fail when evaluated across different datasets. Even with different training strategies such as stable learning \cite{stable}, multi-task learning \cite{mtt,mtt2,mtt3}, or knowledge distillation \cite{distill, distilldmm}, the improvement in generalization is minimal compared to approaches that expand training data or use pretrained models.

Pretrained models like wav2vec 2.0 \cite{wav2vec}, WavLM \cite{wavlm}, and audioMAE \cite{maskedae} prove to be effective front-end feature extractors for FAD, providing enhanced detection performance and improved generalization across different datasets.
There are several researches on how to effectively utilize pretrained models.  
AudioMAE includes an encoder-decoder structure designed for reconstruction tasks during pre-training. This reconstruction process can reveal imperfections in fake audio, making it useful for extracting features that can be applied to FAD \cite{xiaopeng}.

For pretrained models with an encoder structure, using the last layer feature alongside fine-tuning the wav2vec 2.0 model with a classifier can significantly enhance detection performance \cite{com3}.
Additionally, freezing the wav2vec 2.0 model and using the fifth layer hidden state feature \cite{xie02}, rather than the last layer, to train the classifier yields better results.
Research suggests that hidden state features from various layers of pretrained models are beneficial for FAD.
As a result, recent research focus more on leveraging multi-layer features from pretrained models to enhance FAD performance.
For instance, Reference \cite{b1} uses only bona-fide audio for feature distillation training, which reduces the model size while enabling the student model to achieve detection performance comparable to that of the teacher model.
\begin{figure*}
    \centering
    \includegraphics[width=1\linewidth]{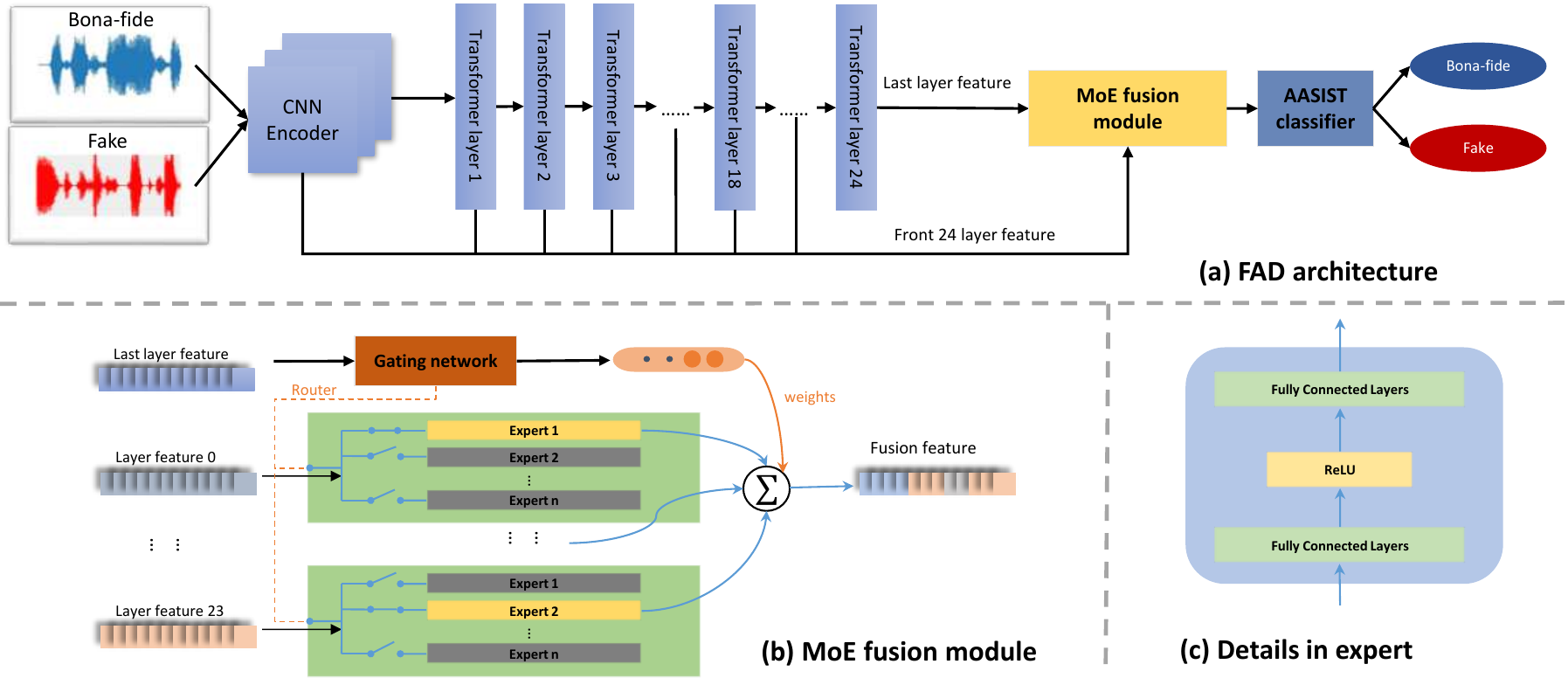}
    \caption{The architecture of FAD architecture (a), details in the MoE fusion module (b) and details in the expert (c).}
    \label{fig:enter-label}
\end{figure*}
In feature fusion methods, the multi-fusion attentive classifier \cite{com4}, the sensitive Layer select module \cite{mmfad} and the attentive merging method \cite{com6} have been proposed to fuse multiple layer features extracted by the wav2vec 2.0 model or WavLM model, all achieving state-of-the-art performance.
The drawback of the aforementioned methods that utilize multi-layer features is that they require fine-tuning the pretrained model, which makes the training process slower and more computationally expensive.

To address the problem mentioned above, in this paper, we propose a module called Mixture of Experts fusion (MoE fusion) that can achieve detection performance competitive with the aforementioned approaches without the need to fine-tune the pretrained model. 
%
The last layer feature of pretrained model is utilized to construct FAD relate knowledge priors in the gating network. 
Then, the expert networks perform dynamic learning of multi-layer features guided by the gating mechanism, achieving sample-adaptive multi-layer feature fusion. 
In addition, we also attempt to explain why freezing the pretrained model yields better results than fine-tuning when using the proposed method. 
Multiple experiments with different Mixture of Experts configurations were conducted, demonstrating that increasing the number of experts or the hidden dimensions of experts improves nothing.


\section{proposed method}
This section briefly introduces the pretrained model used for feature extraction and details the proposed Mixture of Experts fusion method.

\label{sec1}

\subsection{wav2vec 2.0 Model}
The wav2vec 2.0 is a self-supervised learning model for speech recognition developed by Facebook AI. 
It consists of two main components: a feature encoder and a context network. 
The feature encoder is a multi-layer convolutional neural network (CNN) that processes raw audio input to produce latent speech representations.
These representations are then fed into the context network, which is a deep Transformer model composed of 24 Transformer layers, capturing long-range dependencies and producing contextualized representations of the audio.
The context network outputs contextualized representations that commonly used for downstream tasks.

In this paper, we explore using the context network output, called the last hidden state, to guide the fusion of the feature encoder output and the hidden features from the front 23 Transformer layers.

\subsection{Mixture of Experts Fusion}
\label{moef}
As shown in Fig.~\ref{fig:enter-label}, we propose a MoE fusion module to dynamically learning the multiple hidden features from different layers in wav2vec 2.0. 
The MoE \cite{moefather} fusion module is consist of a gating network $G_{all}$ and a set of $N = n \times 24 $ expert networks $\{E_1 , ..., E_{N} \}$.
The structure of each expert network is 2 Fully Connected Layers and 1 ReLU activation layers.
In the MoE fusion module, we flatten the feature $F \in \mathbb{R}^{B \times T \times S}$ to $\hat{F} \in \mathbb{R}^{D \times S}$.  
B represents the batch size. 
T and S denote the output feature dimensions of wav2vec 2.0, which are 201 and 1024, respectively.
The gating network $G_{all}$ take the last hidden state $\hat{F}_{l}^{24}$ as input and produces the probability of it with respective to $n \times 24 $ experts.
The formalization of the gate network is as follows,
\begin{align}
G_{all}(\hat{F}_{l24}) = softmax(TopK(\hat{F}_{l}^{24} \cdot W))
\end{align}
where $W  \in \mathbb{R}^{S \times N}$ is a learnable weight matrix and the TopK outputs are normalized via $softmax$ function.

To achieve specialized learning of different hidden features of different layers in wav2vec 2.0, we firstly flatten each of the hidden features from the front 23 Transformer layers ${F}_{l}^{p}$, where $p \in \{1,...,23\}$, and  the feature encoder output ${F}_{l}^{0}$ to the same shape as the $\hat{F}_{l}^{24}$ ,then pass them into one set of $n$ expert networks $E_n^i$, where $i \in \{0,1,...,23\}$.
There are a total of 24 sets of $n$ expert networks, and thus every $n$ experts that takes each 24 features from different layers as input are non-overlapping \cite{orig}.
Every $n$ expert networks produce their own output $E_n^i(\hat{F}_{l}^{i})$, where $i \in \{0,1,...,23\}$.
The final output $y_{MoEF}$ of the MoE fusion module is calculated as follows,
\begin{equation}
\begin{split}
y_{MoEF}^i = \sum_{j=1}^{n}  G_{all}(\hat{F}_{l}^{24}) E_j^i(\hat{F}_{l}^i) , \\
i = 0,1,...,23 
\label{eq}
\end{split}
\end{equation}

Then the AASIST classifier is applied to obtain the final prediction $Pred$ as follows,
\begin{equation}
\begin{split}
Pred = AASIST(unflatten(concat(y_{MoEF}^i))), \\
i = 0,1,...,23 
\label{eq}
\end{split}
\end{equation}
where the $concat$ means concatenation operation and the $unflatten$ means to revert previously flattened structure back to its original form.

\section{experiments and results}
\label{sec2}

\subsection{Datasets and Metrics}
We utilize the training set of the logical access track in ASVSpoof 2019 dataset \cite{spoof19} (19LA) for training. 
To evaluate the generalization ability of the proposed method, the models are evaluated across multiple evaluation sets of 19LA, logical access track and deepfake track in ASVSpoof 2021 \cite{spoof21} (namely 21LA and 21DF) and InTheWild (ITW) dataset \cite{itw}.
The 21LA dataset consists of utterances transmitted through different channels. 
In contrast, the 21DF dataset comprises approximately 600,000 utterances that have undergone processing with various lossy codecs commonly employed in media storage. 
The 21DF dataset plays a significant role in validating generalization capabilities. 
Both 21LA and 21DF datasets feature a hidden (hid) track where non-speech portions have been removed.

The equal error rate (EER) is used as the evaluation metric. 
The lower the EERs, the better the generalization ability of the models.

\subsection{Implementation Details}
We truncate or repeat all audio samples to a fixed length of approximately 4 seconds (64600 sample points) and do not trim any non-speech portions of training dataset. 

For all experiments, Data augmentation method we use is Rawboost \cite{rb} algorithm 3. 
AdamW optimizer with $\beta$ = [0.9, 0.999] is used for training. 
Cosine schedule with 3 warm-up steps is used to accelerate convergence gradually and we set batch size to 4.
The initial learning rate is set to $10^{-6}$ or $10^{-5}$ when fine-tuning or freezing the wav2vec 2.0 pretrained model during training. 
The training duration is set to 50 epochs, incorporating early stopping technology with a patience of 3. 
Patience indicates that when the training loss does not improve for three epochs, the training will be stopped early, and the model with the lowest training loss is chosen for evaluation.

\subsection{Results and Analysis}
\label{suboercom}
\subsubsection{Performance Comparison}
\label{suboercompf}

\begin{table}[htbp] 
\caption{Performance Comparison\\ EER comparison with other FAD methods using pretrained model in 21LA and 21DF evaluation sets.  }
\renewcommand{\arraystretch}{1.5} 
\begin{center}
\begin{tabular}{llll}
\hline
\multirow{2}{*}{Model} & \multirow{2}{*}{Fine-tune} & \multicolumn{2}{c}{EER(\%)} \\
                                     &  & \multicolumn{1}{c}{21LA} & \multicolumn{1}{c}{21DF}          \\ \hline
RawNet2 \cite{com0}          & -  & {9.50}                   & {22.38}
\\ \hline
wav2vec 2.0 \& LLGF  \cite{com1}       & Yes    & {7.18}                   & {5.44}
\\ \hline
wav2vec 2.0 \& FC \& ASP   \cite{com2}                  & Yes  & 3.54                     & 4.98                              \\ \hline
wav2vec 2.0 \& LGF    \cite{com1}           & Yes  & 6.53                     & 4.75                              \\ \hline
12L-WavLM-Large AttM-LSTM    \cite{com6}           & Yes  & 3.50                     & 3.19                              \\ \hline
10L-WavLM-Large LinM-LSTM    \cite{com6}          & Yes   & 4.52                     & 4.37                              \\ \hline
wav2vec 2.0 \& Rawnet \& ASSIST \cite{com3}  & Yes  & 4.11                     & 2.85                              \\ \hline
wav2vec 2.0 \& AASIST2   \cite{com5}                       & Yes   & \textbf{1.61}                     & 2.77                              \\ \hline
WavLM \& MFA   \cite{com4}                      & Yes    & 5.08                     & 2.56                              \\ \hline
WavLM \& MFA   \cite{com4}                      & No    & -                     & 5.79                              \\ \hline
ensembling model   \cite{com7}                     & No     & 2.31                     & 5.59                              \\ \hline
\textbf{Ours}                             & No     & {2.96} & {\textbf{2.54}} \\ \hline
\end{tabular}
\end{center}
\label{comparison}
\end{table}
Table \ref{comparison} presents the EER results comparison of proposed method and those of other FAD methods that use self-supervised model as feature extractor on 21LA and 21DF evaluation sets.
As is shown, our proposed method using MoE fusion has the compatitive performance on 21LA and best performance on 21DF evaluation set.
Although the results do not show a significant lead, they still demonstrate the effectiveness of the proposed method.

\subsubsection{Ablation Study}
\label{ablation}
\begin{table}[htbp]
\renewcommand{\arraystretch}{1.5} 
\begin{center}
\caption{ 
ablation study \\ 
w/ and w/o means with and without. Freeze indicates that weights in wav2vec 2.0 is fixed during training. Fine-tune means that wav2vec 2.0 and classifier are jointly training. }
\begin{tabular}{lcccc}
\hline
\multirow{2}{*}{Ablation  \textit{(Param)}} & \multicolumn{4}{c}{EER(\%)}          \\ \cline{2-5} 
                          & 19LA & 21LA/hid & 21DF/hid & ITW \\ \hline
\begin{tabular}[c]{@{}l@{}}freeze wav2vec 2.0\\ w/ MoE fusion \textit{(25.92M)} \end{tabular}    & \textbf{0.74} & \textbf{2.96/10.86} & \textbf{2.54/7.32} & 12.48         \\ \hline
\begin{tabular}[c]{@{}l@{}}fine-tune wav2vec 2.0 \\ w/ MoE fusion \textit{(341.35M)}\end{tabular} & 0.89          & 4.25/16.33          & 3.75/13.28         & \textbf{9.17} \\ \hline
\begin{tabular}[c]{@{}l@{}}freeze wav2vec 2.0\\ w/o MoE fusion \textit{(0.44M)}\end{tabular}   & 1.93          & 9.61/20.10          & 7.03/17.83         & 23.22         \\ \hline

\end{tabular}
\label{abla}
\end{center}
\end{table}
In ablation study, we conducted two sets of experiments to demonstrate the effectiveness of the proposed method. 
The experimental results under the setting of "without MoE fusion and fine-tuning the pretrained model" were presented in the previous subsection \ref{suboercompf},  therefore, they will not be discussed in this subsection.
\begin{itemize}
    \item \textbf{Setting 1}: with MoE fusion and fine-tuning the pretrained model.
    \item \textbf{Setting 2}: without MoE fusion, freezing the pretrained model.
\end{itemize}

Experimental results in the Table \ref{abla} show that the Setting 2 leads to a significant decrease in model performance. 
On the other hand, the performance of the Setting 1 is better than the Setting 2 and worse than using the proposed method and freeze the pretrained model. 
What worth mentioning is that fine-tuning the pretrained model performs worse than freezing the pretrained model when using the proposed method for fusing features.
Regarding the reason behind this phenomenon, we tend to explain that when freezing the pretrained model, the whole model do not need to care about the layer features extracted by the pretrained model so that the MoE fusion module can focus on how to integrate layer features and further extract features relevant to FAD for classification.
However, when randomly initializing the classifier and MoE fusion module and then jointly fine-tuning them with the pretrained model transforms the training process into a bi-level optimization problem, which may Lead to difficulties in convergence of the optimization process and make it prone to local optima.

Additionally, the number of trainable parameters in the model is also recorded in the Table \ref{abla}, which highlights that the proposed method's trainable parameters are significantly fewer than those required for fine-tuning a pre-trained model. 
Moreover, the layer features can be extracted from the pre-trained model in advance based on the training dataset, further reducing training time.


\subsubsection{Mixture of Experts Configuration}

In the basic framework of the MoE fusion module, there are three configurable parameters: \textbf{Top-K, the number of experts, and the hidden dimension in experts}. 
The Top-K refers to the number of experts activated for processing each input, used to control the sparsity of the model and balance computational efficiency with model performance. 
The number of Experts refers to the total number of specialized sub-networks or "experts" in the MoE fusion module. 
The hidden dimension in experts refers to the intermediate dimension of the linear layers in the expert structure, used to control the complexity of the experts.
\begin{table}[htbp]
\caption{Mixture of Experts Configuration\\
TopK / Number of Experts / Hidden Dimension in experts configuration in mixture of experts. 
$ft$ indicates that the pretrained model is finetuned using the same configuration as previous row.
}

\renewcommand{\arraystretch}{1.5} 
\begin{center}
\begin{tabular}{lclll}
\hline
\multirow{2}{*}{Configuration} & \multicolumn{4}{c}{EER(\%)}                                              \\ \cline{2-5} 
                               & \multicolumn{1}{l}{19LA}          & \multicolumn{1}{l}{21LA/hid}            & \multicolumn{1}{l}{21DF/hid}           & \multicolumn{1}{l}{ITW}     \\ \hline
2/4/128                        & 0.74          & \textbf{2.96/10.86} & \textbf{2.54/7.32} & 12.48         \\
$ft$                             & 0.89          & 4.25/16.33          & 3.75/13.28         & \textbf{9.17} \\ \hline
2/6/128                        & \textbf{0.43} & 4.83/18.42          & 3.93/11.32         & 19.00         \\
$ft$                             & 1.41          & 7.22/21.44          & 5.39/16.58         & 13.93         \\ \hline
2/8/128                        & 0.44          & 3.89/15.82          & 3.47/9.28          & 16.62         \\
$ft$                             & 1.04          & 16.23/35.15         & 8.10/27.42         & 10.59         \\ \hline
1/4/128                        & 77.05         & 75.36/43.73         & 66.79/42.78        & 55.69         \\
$ft$                             & 83.04         & 81.87/45.76         & 69.08/38.07        & 40.54         \\ \hline
3/4/128                        & 1.60          & 5.16/13.97          & 3.16/9.70          & 18.21         \\ \hline
2/4/512                        & 1.04          & 5.26/14.75          & 3.47/10.00         & 15.78         \\ \hline
2/4/1024                       & 1.17          & 4.50/19.24          & 3.28/12.72         & 16.50         \\ \hline
\end{tabular}
\end{center}
\label{moe}
\end{table}
In this paper, three sets of comparative experiments are conducted to analyze the effects of these hyperparameters mentioned above. 
The models with top-k, the number of experts per feature ({n} mentioned in Sec.\ref{moef}), and hidden dimensions set to 2, 4, and 128 are used as the baseline. 

When setting the number of experts per layer feature to 6 or 8, there are improvements in both EER results on the 19LA evaluation set, but the EER results on the other three evaluation sets decline compared to the baseline model. 
This demonstrates that increasing the number of experts does not necessarily improve the model's generalization performance; in fact, it may lead to overfitting.

When reducing and increasing the Top-K also leads to decreased performance, especially when Top-K is set to 1, indicating that excessive sparsity is not helpful for FAD.

When the hidden dimension is set to 512 and 1024, performance also declines. This suggests that each layer of wav2vec 2.0 features contains redundant information that is irrelevant to FAD. 
Therefore, performing dimensionality reduction within the experts is more beneficial for FAD.

In addition, this subsection also conducted experiments with three sets of fine-tuned pretrained models. The experimental results are consistent with those in the previous Sec.\ref{ablation}, which indirectly demonstrates the effectiveness of the proposed method and the accuracy of the explanation.

\section{conclusion}
\label{sec3}
In this paper, we propose a feature fusion technique based on mixture of experts, named MoE fusion, and address the problem that previous feature fusion methods required fine-tuning pretrained models, which led to excessively long training duration. 
Experimental results conducted on the ASVspoof2019 and ASVspoof2021 datasets show that freezing and using the last layer feature of wav2vec 2.0 as gating network input to guide the mixture of experts for combining 24 hidden features from different layers of wav2vec 2.0 can further improve FAD performance and generalization. 
Additionally, we conducted multiple ablation and comparison experiments to demonstrate the effectiveness of the proposed method and analyze the reasons why fine-tuning performs better than freezing the pretrained model.
In summary, the proposed MoE fusion method effectively integrates the FAD-related components from the representations extracted by different layers of the pretrained model. 

Future work will explore how to use fewer layers of pretrained model and less parameters to achieve feature fusion while enhancing the model's FAD performance.


\bibliographystyle{IEEEtran}\
\bibliography{main}

\end{document}